\documentclass[pdflatex,sn-mathphys-num]{sn-jnl}% Math and Physical Sciences Numbered Reference Style
\usepackage{graphicx}%
\usepackage{multirow}%
\usepackage{amsmath,amssymb,amsfonts}%
\usepackage{amsthm}%
\usepackage{mathrsfs}%
\usepackage[title]{appendix}%
\usepackage{xcolor}%
\usepackage{textcomp}%
\usepackage{manyfoot}%
\usepackage{booktabs}%
\usepackage{hhline}%
\usepackage{array}
\usepackage{algorithm}%
\usepackage{algorithmicx}%
\usepackage{algpseudocode}%
\usepackage{listings}%

\theoremstyle{thmstyleone}%
\theoremstyle{thmstyletwo}%

\theoremstyle{thmstylethree}%

\begin{document}

\title[Article Title]{Mass conservation analysis of extrusion-based 3D printing simulations based on the level-set method}


\author*[1]{\fnm{Carlos J.G.} \sur{Rojas}}\email{c.j.gonzalez.rojas@tue.nl}

\author[2,3]{\fnm{Md. Tusher} \sur{Mollah}}\email{mtumo@mek.dtu.dk, tusher@iut-dhaka.edu}

 \author[1]{\fnm{César A.} \sur{Gómez-Pérez}}\email{c.a.gomez.perez@tue.nl}

\author[1]{\fnm{Leyla} \sur{Özkan}}\email{l.ozkan@tue.nl}


 \affil*[1]{\orgdiv{Control Systems Group }, \orgname{Eindhoven University of Technology}, \country{Netherlands}}

 \affil[2]{\orgdiv{Department of Civil and Mechanical Engineering}, \orgname{Technical University of Denmark}, , \country{Denmark}}

 \affil[3]{\orgdiv{Department of Natural Sciences}, \orgname{Islamic University of Technology},  \country{Bangladesh}}


\abstract{Accurate numerical simulation of material extrusion additive manufacturing requires reliable tracking of evolving material interfaces while preserving mass conservation. Inaccurate mass conservation can lead to significant discrepancies between simulated and deposited strand geometries, undermining the predictive capability of the model. In this work, we investigate the mass conservation performance of the conservative level-set (CLS) method in extrusion-based 3D printing simulations. A systematic parametric study is conducted to quantify the influence of the interface thickness and reinitialization parameters on mass conservation, using the steady-state cross-sectional area of deposited strands as a quantitative metric. Simulated cross-sections are compared against reference values obtained from analytical mass balance relations. The results show that reducing both the interface thickness and the reinitialization parameter improves mass conservation accuracy, although diminishing returns and increased computational cost are observed beyond certain thresholds. In addition, appropriate tuning of the interface thickness can relax mesh refinement requirements while maintaining acceptable accuracy. The proposed parameter selection strategy is validated across a range of printing conditions, materials, and nozzle geometries, including multilayer deposition of viscoplastic fluids. The simulations show reasonable agreement with experimentally validated data from the literature, confirming that careful CLS parameter tuning enables accurate and computationally efficient prediction of strand geometry in extrusion-based 3D printing.}


\maketitle

\section{Introduction}\label{sec1}

{Extrusion-based 3D printing, also known as Material Extrusion Additive Manufacturing (MEX-AM)}, has become widely adopted due to its ability to process a diverse set of materials, including bioinks, ceramics, food, and concrete, across scales ranging from micrometers to centimeters \citep{6_altiparmak_extrusion-based_2022}. Despite this versatility, achieving a reliable and consistent process remains challenging, particularly when experimental iteration is costly or limited \citep{7_naghieh_printabilitykey_2021}. As a result, numerical simulation has been recognized as an effective means of investigating how material properties (e.g., viscosity, yield stress, and surface tension) and processing conditions  (e.g., extrusion pressure, nozzle velocity, curing, and layer height) affect strand formation, surface quality, and interlayer bonding \citep{3_balta_numerical_2022, 11_mollah_stability_2021, Lukhi2025}.

Simulating MEX-AM requires multiphysics modeling of a two-phase flow, in which the extruded material interacts with the surrounding air and the fluid interface evolves over time. A consistent definition of the interface is therefore essential for enforcing mass conservation and managing discontinuities, making its numerical treatment a critical element of the solution strategy \citep{5_Mirjalili2017_interface_capturing}. Choosing the right approach for interface modeling is therefore a key step in ensuring accurate and reliable simulation results. Depending on how the computational mesh is used, interface-handling techniques can be broadly categorized into interface-capturing and interface-tracking methods, with the former adopting a Eulerian description on a fixed mesh and the latter relying on a Lagrangian or hybrid approach that follows the interface motion \citep{elgeti_deforming_2016}. In 3D printing simulations, the most popular solutions are based on interface-capturing methods, including the volume of fluid (VOF) \citep{10_spangenberg_numerical_2021,11_mollah_stability_2021}, the coupled level-set with VOF \citep{1_comminal_numerical_2018, 9_serdeczny_numerical_2019}  and the conservative level-set (CLS) \citep{3_balta_numerical_2022,8_bakrani_balani_investigation_2023}. The primary distinction between these methods lies in how they represent the interface. VOF methods employ a sharp interface, which ensures good mass conservation but requires more complex numerical treatment to handle discontinuities. In contrast, the CLS method uses a continuous, diffuse interface that is simpler to implement but requires careful parameter tuning to maintain mass conservation \citep{5_Mirjalili2017_interface_capturing}.

Among the interface-capturing methods, the conservative level-set (CLS) approach is particularly appealing for MEX-AM simulations because it combines a smooth interface representation with improved conservation properties. The CLS approach was originally proposed to retain the simplicity of the classical level-set method while improving its mass conservation capabilities \citep{12_olsson_conservative_2005}. This formulation employs a smooth level-set function along with a specific advection scheme to enhance conservation. However, a reinitialization step is required to maintain a constant interface thickness, which can affect convergence and, in complex flows or long simulations, lead to error accumulation \citep{15_desjardins_accurate_2008,14_mccaslin_localized_2014,16_chiodi_reformulation_2017}. To address this limitation, a more computationally efficient variant \citep{18_chiu_conservative_2011} combines the advection and reinitialization steps into a single equation, reducing computational demand, although it requires careful selection of the interface thickness and the reinitialization parameters. While the CLS method offers distinct advantages, its predictive accuracy depends critically on parameter selection. Therefore, accurate selection of the CLS parameters is crucial in MEX-AM simulations, as it could directly influence predictions of strand cross-section, surface roughness, and interlayer adhesion.

Although previous studies have evaluated the influence of process settings on strand geometries via simulations validated with experiments \citep{1_comminal_numerical_2018, 8_bakrani_balani_investigation_2023, 3_balta_numerical_2022}, systematic analysis of parametric effects, especially regarding mass conservation, has been lacking. Given that mass conservation can directly affect the quality of predictions, understanding the CLS parameters' influence is essential for accurate simulation. In this work, 
we investigate the influence of CLS parameters on mass conservation and strand formation, providing quantitative metrics to assess simulation accuracy. We employ the conservative level-set method to simulate multilayer MEX-AM, extending the formulation with a moving-mesh approach to accommodate vertical motion of the substrate. The simulation results are validated against experimental and simulation data for single and multilayer deposition with different materials and nozzle geometries. 

The remainder of this article is organized as follows. In Section II, we present all the relevant details for the modeling and simulation. Section III describes the methodology of the parametric analysis and the single and multilayer study cases. In Section IV, we discuss the effects of the level-set parameters and evaluate the parameter selection for multiple case studies. Finally, the conclusions of this work are given in Section V.

\section{Model Description}\label{sec2}

The model used in this study is based on a one-fluid formulation, in which the dynamics of the flow is governed by a single set of Navier-Stokes equations and the interface between phases is captured using the level-set method. Within this framework, the single-step CLS method tracks the evolving interface, with coupling achieved through the advection velocity and averaging of phase properties. In-plane motion is imposed through a boundary condition; however, multilayer printing introduces relative vertical motion, which requires updating one of the domain boundaries. To accommodate this, the formulation is extended with a moving-mesh approach (arbitrary Lagrangian–Eulerian), allowing the material frame to be applied locally while maintaining accurate interface tracking and conservation.
\begin{figure}[!htbp]
  \centering    \includegraphics[width=1.0\textwidth]{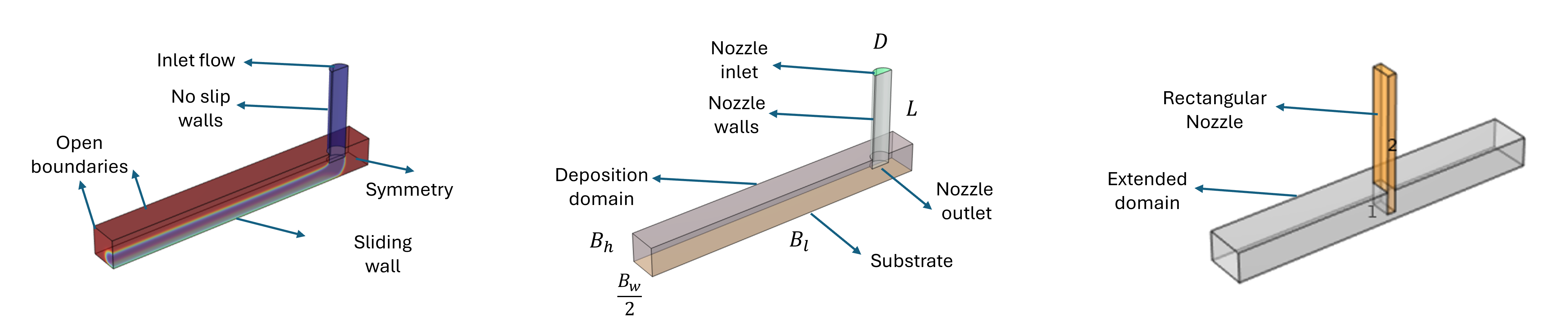}
  \caption{a) Boundary conditions. b) Single-layer simulation domain. c) Multilayer simulation domain}
  \label{fig:1}
\end{figure}

\subsection{Single-layer simulations} 
 The domain for single-layer simulations was defined according to the work in \cite{2_serdeczny_experimental_2018} and a three-dimensional representation of the domain considering the $xz$ symmetry is depicted in Fig  \ref{fig:1}b. The nozzle was modeled as a cylinder with diameter $D$ and length $L$, and the distance between the tip of the nozzle and the bed is $\Delta Z$. The deposition domain was represented by a rectangular block, with cross-sectional dimensions $B_w \times B_h$ and length $B_l$.  The effects of gravity were considered in the simulations %and, for simplicity, the printing material was represented by a Newtonian constitutive model with viscosity $\mu$. 
 and all properties and dimensions used in the single-layer simulations are given in Table 1.

\begin{table}
\centering
\caption{ Single layer - Simulation Parameters}
\begin{tabular}{|l|c|c|c|}
\hline
\textbf{Parameter} & \textbf{Nomenclature} & \textbf{Values} & \textbf{Units} \\
\hline
$D$ & Nozzle Diameter & 0.4 & $mm$ \\
$L$ & Nozzle Length & 2 & $mm$ \\
$\Delta Z$ & Nozzle to bed distance & \{0.24, 0.32, 0.4\} & $mm$ \\
$B_w$ & Block width & 1.2 & $mm$ \\
$B_h$ & Block height & 0.6 & $mm$ \\
$B_l$ & Block Length & 6 & $mm$ \\
$v_{p}$ &Plunger speed  & 20 & $mm/s$ \\
$V_{pl}$ & Inlet flow & 1.257 & $mm^3/s$ \\
$\bar{v}_x$ & Printing speed & \{10, 15, 20\} & mm/s \\
$\rho$ & Density of the material & 1.0 & g/cm$^3$ \\
$\mu_s$ & Viscosity of the material & 1000 & $Pa \cdot s$ \\
$g$ & gravity acceleration & $-9.81$ & $m/s^2$ \\
\hline
\end{tabular}
\end{table}

\subsection{Multilayer simulations}
{
The multilayer simulations followed the printing settings reported by \cite{el_abbaoui_3d_2024}, \cite{10_spangenberg_numerical_2021} and \cite{co_comminal_modelling_2020} for the deposition of concrete using a cylindrical nozzle, and by \cite{Mollah2025RapidCuringCFD} for the deposition of metakaolin using a rectangular nozzle. The nozzle was placed in the center of the deposition domain, and the length $B_l$ was increased to accommodate the horizontal motion in both directions. The simulation domain for the rectangular nozzle is shown in Figure  \ref{fig:1}c, and the domain for the cylindrical nozzle followed the same configuration.  The smoothing of the moving mesh is based on the Yeoh method with a stiffening factor $C_2=100$. The boundaries defining the nozzle were fixed, and the substrate wall had a prescribed horizontal speed (printing speed) and a normal mesh velocity (vertical speed), as shown in Figure 2. 
\begin{figure}[!htbp]
  \centering  
\includegraphics[width=0.75\textwidth]{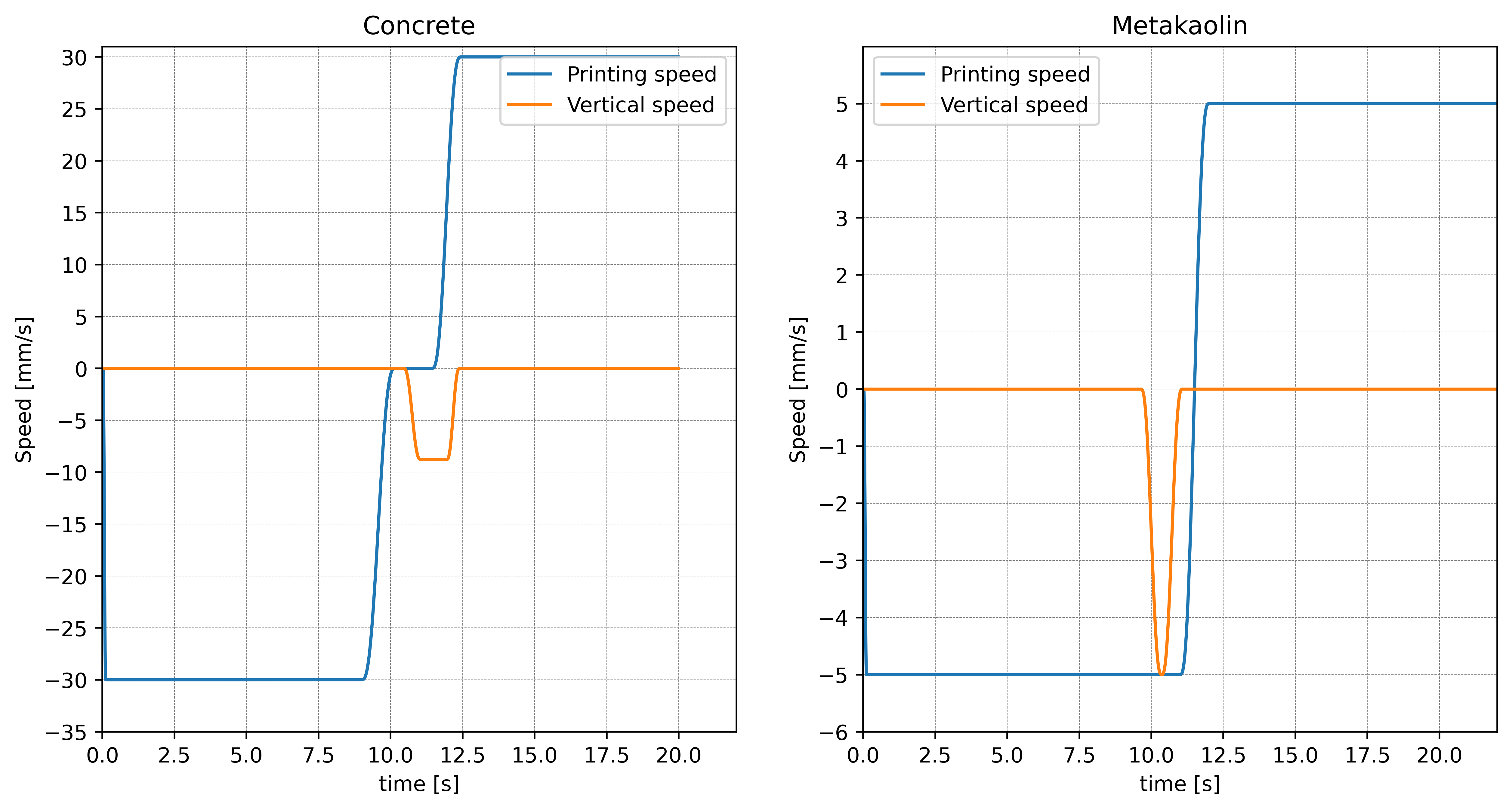}
  \caption{Printing and vertical speeds a) Concrete b) Metakaolin}
  \label{fig:2}
\end{figure}
The cylindrical nozzle has a diameter $ D = 25$ $mm$ and the rectangular nozzle has a cross section of $20 \times 4$ $mm^2$. All the properties and dimensions used for the multilayer simulations are given in Table 2.
\begin{table}[h!]
\centering
\caption{Multilayer - Simulation Parameters}
\begin{tabular}{|l|c|c|c|c|}
\hline
\textbf{Parameter} & \textbf{Nomenclature} & \textbf{Values (Concrete)} & \textbf{Values (Metakaolin)} & \textbf{Units} \\
\hline
$L$ & Nozzle Length & 150 & 67 & $mm$ \\
$\Delta Z$ & Nozzle to bed distance & 12.5 & 3.6 & $mm$ \\
$B_w$ & Block width & 80 & 40 & $mm$ \\
$B_h$ & Block height & 50 & 14.4 & $mm$ \\
$B_l$ & Block Length & 630  & 150 & $mm$ \\
$V_{pl}$ & Inlet flow & 8240 & 180 & $mm^3/s$ \\
$\rho$ & Density of the material & 2.1 & 1.63 & $g/cm^3$ \\
$\mu_p$ & Plastic Viscosity & 7.5 & 0.4746 & $Pa \cdot s$ \\
$\tau_y$ & Yield stress  & 630 & 41.69 & $Pa$ \\
$m_p$ & Bingham regularization  & 100 & 100 & $s$ \\

\hline
\end{tabular}
\end{table}

}
\subsection{Fluid flow equations}
\label{subsection_Modeling}

The equations of fluid motion for the extrusion and deposition of material provide fundamental macroscopic variables such as velocity and pressure. Considering an isothermal and incompressible flow, the Navier-Stokes equations describe the transport of mass and momentum:
\begin{equation}
\label{eq:ns1}
\mathbf{\nabla} \cdot \mathbf{v} = 0
\end{equation}
\begin{equation}
\label{eq:ns2}
\rho \left(\frac{\partial \mathbf{v}}{\partial t} + (\mathbf{v} \cdot  \mathbf{\nabla}) \mathbf{v}\right) = \mathbf{\nabla} \cdot\left( -p\mathbf{I}  +   \mathbf{\sigma}\right) + { \mathbf{F}_g + \mathbf{F}_s}
\end{equation}
where $\rho$ denotes density, $\mathbf{v}$ 
velocity, $p$ pressure,  $\sigma$ the deviatoric stress tensor, { $\mathbf{F}_g$ the body force term from gravity and $\mathbf{F}_s$  the force due to surface tension.}
A generalized stress tensor is defined as:
\begin{equation}
\label{eq:ns3}
\sigma = 2\mu \mathbf{D}
\end{equation}
%\begin{equation}
%\label{eq:ns4}
%\dot{\gamma} =  \left(2\mathbf{D}:\mathbf{D}\right)^\frac{1}{2}
%\end{equation}
\begin{equation}
\label{eq:ns4}
\mathbf{D} =  \left(\frac{\mathbf{\nabla} \mathbf{v} + (\mathbf{\nabla} \mathbf{v})^\intercal}{2}\right)
\end{equation}
here $\mu$ is the viscosity and $\mathbf{D}$ is the strain-rate tensor. {The viscosity of the ink is governed by its composition and microstructure, together with processing conditions such as shear rate, temperature, and printing parameters, and is characterized using rheological tests \citep{wilms_formulation_2021, BakraniBalani2025, Mishra2022}.
{
As reported in \cite{20_Serdeczny2018_shear_thinning}, a Newtonian model can be employed when the primary objective is to predict the cross-sectional shape of the deposited strand. In contrast, more complex models can be adopted to capture more complex effects, such as elasticity and yield stress, which improve the prediction of printing forces \citep{serdeczny_viscoelastic_2022}.  In this study,  a Newtonian constitutive model with viscosity $\mu =\mu_s$ is used for the single-layer simulations, and the printing materials for the multilayer simulations are modeled using the Bingham-Papanastasiou constitutive model:
\begin{equation}
\mu_m =\mu_p+\frac{\tau_y}{\dot{\gamma}}(1-\exp(-m_p\dot{\gamma} )) 
\end{equation}
In this case $\mu =\mu_m$ and the model is parameterized by the plastic viscosity $\mu_p$, the yield stress $\tau_y$, the shear rate $\dot{\gamma}$, and the regularization parameter $m_p$.}

\subsection{Conservative level-set method}
{
The conservative level-set method proposed in \cite{12_olsson_conservative_2005} introduces a continuous scalar field $\phi \in [0,1]$ that defines the interface between two fluids as the contour $\phi = 0.5$. The evolution of the interface between the two phases requires the advection of $\phi$ based on the velocity $\mathbf{v}$ and is given by the following conservative form:
%\eqref{eq:ns5} and \eqref{eq:ns6}) as follows:
\begin{equation}
\label{eq:ns5}
\frac{\partial {\phi}}{\partial t} +  \mathbf{\nabla} \cdot (\mathbf{v}\phi) = 0
\end{equation}
This expression has a simple structure and is especially suitable for simulations of incompressible flow. Nevertheless, as shown in \cite{13_olsson_conservative_2007}, it is necessary to stabilize Eq. \eqref{eq:ns5} because numerical perturbations are not damped and create distorted shapes during the solution. \cite{13_olsson_conservative_2007} formulated a stabilized advection based on an additional term that depends on a reinitialization parameter { ${\gamma}$ } and a parameter controlling the thickness of the interface $\epsilon$,
\begin{equation}
\label{eq:ns6}
\frac{\partial {\phi}}{\partial t} +  \mathbf{\nabla} \cdot (\mathbf{v}\phi) = \gamma \mathbf{\nabla} \cdot\left(  \epsilon (\nabla \phi \cdot \hat{n})\hat{n}  - \phi(1-\phi) \hat{n}\right), \qquad\text{where} \quad \hat{n} = \frac{\nabla \phi}{|\nabla \phi|}.
\end{equation}
However, the solution of this equation is usually divided into advection and stabilization steps for numerical reasons. In this work, we use the single-step CLS method \citep{18_chiu_conservative_2011}, which does not require a separate stabilization step. The advection of the level-set function is based on the following expression:
\begin{equation}
\label{eq:ns7}
\frac{\partial {\phi}}{\partial t} +  \mathbf{\nabla} \cdot (\mathbf{v}\phi) = \gamma \mathbf{\nabla} \cdot\left(  \epsilon \nabla \phi - \phi(1-\phi) \frac{\nabla \phi}{|\nabla \phi|}\right)
\end{equation}
The only difference between the stabilized advection (Eq. \eqref{eq:ns6}) and the single-step level-set evolution (Eq. \eqref{eq:ns7}) is the projection of the gradient $\nabla \phi$ along the normal direction $\hat{n}$. Since the single-step CLS method requires the solution of a single PDE, its solution is computationally more efficient than the original conservative level-set method. Furthermore, if Eq. \eqref{eq:ns7}  is solved using conservative numerical schemes and the level-set parameters $\epsilon$ and $\gamma$ are properly tuned, then the mass of the fluids should be conserved.

A variety of empirical and analytical guidelines have been proposed for tuning the level-set parameters. \cite{13_olsson_conservative_2007} showed that the effect of numerical perturbations on the solution evolves on a time scale of { $t \sim \epsilon / \gamma$}. Based on this timescale, it is assumed that if { $\epsilon  / \gamma \ll 1$}, perturbation effects are significantly faster than the advection dynamics. The interface thickness $\epsilon$ is typically chosen to be on the order of the maximum element size of the mesh \citep{13_olsson_conservative_2007, 18_chiu_conservative_2011}. Regarding the reinitialization parameter, \cite{13_olsson_conservative_2007} recommended using { $\gamma = 1$} for constant velocities (when $\epsilon$ is small), or setting { $\gamma$} according to the velocity gradients at the interface. \cite{18_chiu_conservative_2011} selected $\gamma$ to match the maximum velocity in the domain. Other studies have proposed criteria for selecting $\epsilon$ and $\gamma$ based on the stability of the numerical solution and the boundedness of the level-set field $\phi$ \citep{19_mirjalili_conservative_2020, 172_jain_conservative_2020}. These parameters have also been shown to constrain the allowable time-step size, potentially increasing the computational cost \citep{17_jain_accurate_2022}. However, these analyses are typically based on uniform meshes and central finite difference schemes, which limits their direct applicability to other discretization methods.}

\subsection{Boundary and coupling conditions}
The boundary conditions of the model are the no-slip condition on the nozzle walls, symmetry around the $xz$ plane, an inlet flow rate on the top of the cylinder, the sliding wall at the bottom of the rectangular block, and open boundaries for the rest of the walls in the deposition domain. The domain and the boundaries considered in a general 3D printing simulation are represented in Figure 1. From an operational perspective, two boundary conditions deserve more attention because they incorporate the manipulated variables of the process as follows:
\begin{equation}
\label{eq:ns8}
V_{pl} = -\int_{\partial \Omega_{inl}} \mathbf{v}\cdot \mathbf{n} \,dS ,
\end{equation}
\begin{equation}
\label{eq:ns9}
\mathbf{v}_{\partial \Omega_{bn}} = \mathbf{v}_{bn}, \quad \mathbf{v}_{bn} = (\bar{v}_x, 0, 0)
\end{equation}
In Eq \ref{eq:ns8}, the normal contributions of $\mathbf{v}$ over the boundary $\partial \Omega_{inl}$ (nozzle inlet) are given by the inlet flow rate $V_{pl}$, and in Eq \ref{eq:ns9}, the velocity on the boundary $\partial \Omega_{bn}$ (substrate) is set to $\bar{v}_x$, the relative printing speed. In terms of control, $\bar{v}_x$ can be directly adjusted using the G code, but $V_{pl}$ is implicitly related to the dynamics of the extrusion system. % When the system is driven by pressure,  a different boundary condition can also be adopted.

The fluid flow and the level-set equations are coupled through the advection velocity, but also through the properties of the phases. The density and viscosity used in the fluid flow equations are defined based on a smooth transition given by the following expressions:
\begin{equation}
\rho = \rho_1 + (\rho_2-\rho_1) H(\phi)
\end{equation}
\begin{equation}
\mu = \mu_1 + (\mu_2-\mu_1) H(\phi)
\end{equation}
where $H$ is a smooth Heaviside function \citep{13_olsson_conservative_2007,18_chiu_conservative_2011}.

\subsection{Moving mesh}
{
Two-phase flow simulations based on the one-fluid formulation require a multiphysics approach to capture the interface between the fluids. The horizontal motion of the strand can be imposed using a sliding wall boundary condition on the substrate. However, a multilayer printing simulation requires a relative vertical displacement of the bed (or the nozzle), which cannot be incorporated by using a kinematic condition at the boundary. Since a purely Eulerian formulation cannot accommodate moving boundaries, a reformulation of the equations is needed. Therefore, we define a moving mesh to handle the change in the domain when the system has the aforementioned vertical motion. This feature is based on an arbitrary Lagrangian-Eulerian (ALE) method where the equations are expressed in a reference frame that follows the motion of the computational mesh. The mesh coordinates are dynamically updated to maintain mesh quality, with the mapping between the mesh and spatial coordinates adjusted as necessary \citep{comsol}}.

\section{Methodology}
\subsection{Simulation settings}
The simulations were implemented in COMSOL Multiphysics\textsuperscript{\tiny\textregistered} v6.2 \citep{comsol}, using the Laminar Flow and level-set interfaces coupled with the Two-Phase Flow level-set multiphysics. A stationary study step was defined to have a consistent initialization of the time-dependent solution, and the boundary conditions were ramped up to avoid numerical issues. All simulations performed in this work were implemented in the HPC cluster of the Eindhoven University of Technology. The jobs run on only one node with one task and 32 CPUs per task.

{\subsection{Parametric analysis of the level-set method}}
{In the present study, we assess the influence of the level-set parameters on the numerical results. As presented in Section 2.2, the level-set parameters affect the numerical accuracy and the computational cost of the CLS solutions. In addition, the analytical relationships available for $\epsilon$ and $\gamma$ are only valid for specific discretization schemes. Therefore, we propose a parametric analysis to isolate the individual effects of $\epsilon$ and $\gamma$ on the mass conservation of 3D printing simulations.
For simplicity, this analysis is done considering single-layer simulations, but is evaluated with multilayer printing scenarios.

}
\subsubsection{Mass conservation analysis}
Due to the high viscosity effects and the negligible influence of the inertial forces, a printed strand quickly reaches a steady state.  Therefore, applying mass conservation (on a suitable control volume) and assuming a constant density, the following relationship holds:
\begin{equation}
V_{pl} = A_f \bar{v}_x
\end{equation}
where $A_f$ is the area of the strand and $\bar{v}_x$ is the printing speed. In addition, the inlet flow rate $V_{pl}$ can be directly related to the cross-sectional area of the nozzle $\pi D^2/4 $ and the speed of the plunger $v_p$, which are settings of the process. We considered $\Delta Z = 0.32 \ mm$ and $\bar{v}_x = 20 \ mm/s$ to investigate the effects of the level-set parameters.  In \cite{1_comminal_numerical_2018,2_serdeczny_experimental_2018}, the authors compared the compactness of the strands obtained in the simulation with idealized cross sections. In contrast, we directly compared the area $A_f$ calculated from the mass conservation in Eq. 13 to the following surface area:
\begin{equation}
A_s = \int_{\Omega} \xi(\phi)\,dA 
\end{equation}
\begin{equation}
\xi (\phi) = \begin{cases}
1, & \text{if } \phi \leq 0.5 \\
0, & \text{otherwise}
\end{cases}
\end{equation}
with $\xi$ evaluated on all elements and $A_s$ calculated on a predefined cut plane for the different parameters of the level-set method considered in this study. For $\epsilon$, we evaluated how the change in thickness around the maximum mesh element size $m_{max}$ affects the cross-sectional area of the fluid. In the case of $\gamma$,  we proposed to use values around the printing speed $\bar{v}_x$, as this captures the most critical speed change. 
At the beginning of the simulation, the fluid takes some time to spread over the bed, and a fixed reinitialization rate can lead to mass conservation errors due to the proximity of the active (moving) and inactive (stationary) interfaces \citep{14_mccaslin_localized_2014}. Therefore, the mass conservation analysis was carried out using cross-sections of the strand  $2 \ \mathrm{mm}$ away from the nozzle axis at $t = 0.5 \mathrm{s}$.

\subsubsection{Mesh convergence analysis}
We performed a simple mesh refinement study to test the convergence of the solution for the settings $\Delta Z = 0.32 \ mm$ and $\bar{v}_x = 20 \ mm/s$. As a starting point for this analysis, we considered the recommended $\epsilon$ in COMSOL:
\begin{equation}
\epsilon_{ref} = \begin{cases}
m_{max}, & \text{if } m_{max}>1.3*m_{min}\\
2*m_{max}, & \text{otherwise}
\end{cases}
\end{equation}
where $m_{max}$ and $m_{min}$ are the maximum and minimum element size around the fluid's interface. We used $\gamma = \{0.02, 1\} \ m/s$, the first value is based on the printing speed defined for this study, and the second is the default value for the reinitialization parameter in COMSOL. The variables used to evaluate the convergence of the simulations were the surface area $A_s$ and the maximum pressure on the nozzle $P_{max}$. With these two variables, we account for the effects of the mesh on both the fluid dynamics and the level-set solutions. For the level-set solution, we used as a reference the strand area $A_f$ given by Eq. 13. Since the mesh size affects $\epsilon$, we also considered the difference between $A_s$ and $A_f$:
\begin{equation}
\Delta A = \frac{|A_s-A_f|}{A_f} \times 100,
\end{equation}
to select a single mesh for the mass conservation study. As a reference for the pressure, we considered the analytical Hagen-Poiseuille equation:
\begin{equation}
\Delta P = \frac{8 \mu L V_{pl}}{\pi R^4}.
\end{equation}
A physics-controlled mesh with a fine element size calibrated for fluid dynamics was defined. In the mesh settings, only the maximum and minimum element sizes were changed during the convergence analysis. The other mesh settings, element growth rate (1.08), curve factor (0.3), and resolution of narrow regions (0.95), were kept constant in all the numerical solutions. 

\section{Results}
{In this section, we present a parametric study to evaluate the accuracy of the conservative level-set formulation as a function of the reinitialization factor and interface thickness. The analysis focuses on assessing mass conservation and examining the resulting strand geometry over a range of level-set parameters selected based on recommendations from the literature. Building upon this framework, we further extend the simulations to multilayer printing scenarios involving Bingham fluids with different rheological properties and nozzle cross-section geometries.}

\subsection{Effects of the level-set parameters in single-layer printing}
This subsection investigates the sensitivity of the simulated strand geometry and mass conservation to variations in the interface thickness and the reinitialization parameters within the conservative level-set framework. We begin by assessing mesh convergence using the ideal strand area derived from the mass balance and the maximum pressure within the domain. This is followed by a systematic set of single-layer deposition tests aimed at quantifying the influence of the level-set parameters on the accuracy of the numerical results. 

\subsubsection{Mesh convergence results}
The values of reference for the mesh convergence test were $A_f = 62.83 \times 10^{-3}  \ [mm^2]$ (for $\bar{v}_x = 20 \ mm/s$) and $\Delta P = 8 $ [MPa]. The convergence results are presented in Table 3. 
\begin{table}[h!]
\centering
\caption{Mesh comparison}
\renewcommand{\arraystretch}{1.2}
\begin{tabular}{|c|c|c|c|c|c|}
\hline
\multicolumn{1}{|c|}{} & \textbf{Number of elements}
& \multicolumn{2}{c|}{$A_s \times 1000 \,[\mathrm{mm}^2]$}
& \multicolumn{2}{c|}{$P_{\max}\,[\mathrm{MPa}]$} \\
\hhline{|~|~|----|}
\multicolumn{1}{|c|}{} &
& $\gamma = 0.02$ & $\gamma = 1$
& $\gamma = 0.02$ & $\gamma = 1$ \\
\hline
\textbf{Mesh 1} & 2\,077\,795 & 59.65 & 40.22 & 8.393 & 8.263 \\
\hline
\textbf{Mesh 2} & 3\,315\,894 & 61.58 & 51.45 & 8.381 & 8.323 \\
\hline
\textbf{Mesh 3} & 4\,957\,174 & 62.36 & 56.22 & 8.377 & 8.347 \\
\hline
\end{tabular}
\end{table}
The maximum pressure in the domain $P_{max}$ converged faster than $A_s$, which required a finer mesh to reach a relatively convergent value. Although $A_s$ was still changing, we did not refine it further because the value obtained for {Mesh 3} is almost the same as the analytical value $A_f$ with less than $1 \%$ error. Similarly, $P_{max}$ exceeded the reference $\Delta P$ by less than $5 \%$.
The effects of $\gamma$ are discussed in the following subsection; however, we already notice a big difference between $A_s$ and the reference area $A_f$ depending on the value of $\gamma$.  Based on the computational time required for $\gamma = 0.02$, we decided to use {Mesh 1} because it took less simulation time compared to {Mesh 3} and still gave an acceptable error of around $5 \%$ in $A_s$.

\subsubsection{Effects of the reinitialization parameter}
\begin{figure}[!htbp]
  \centering  
  \includegraphics[width=0.5\textwidth]{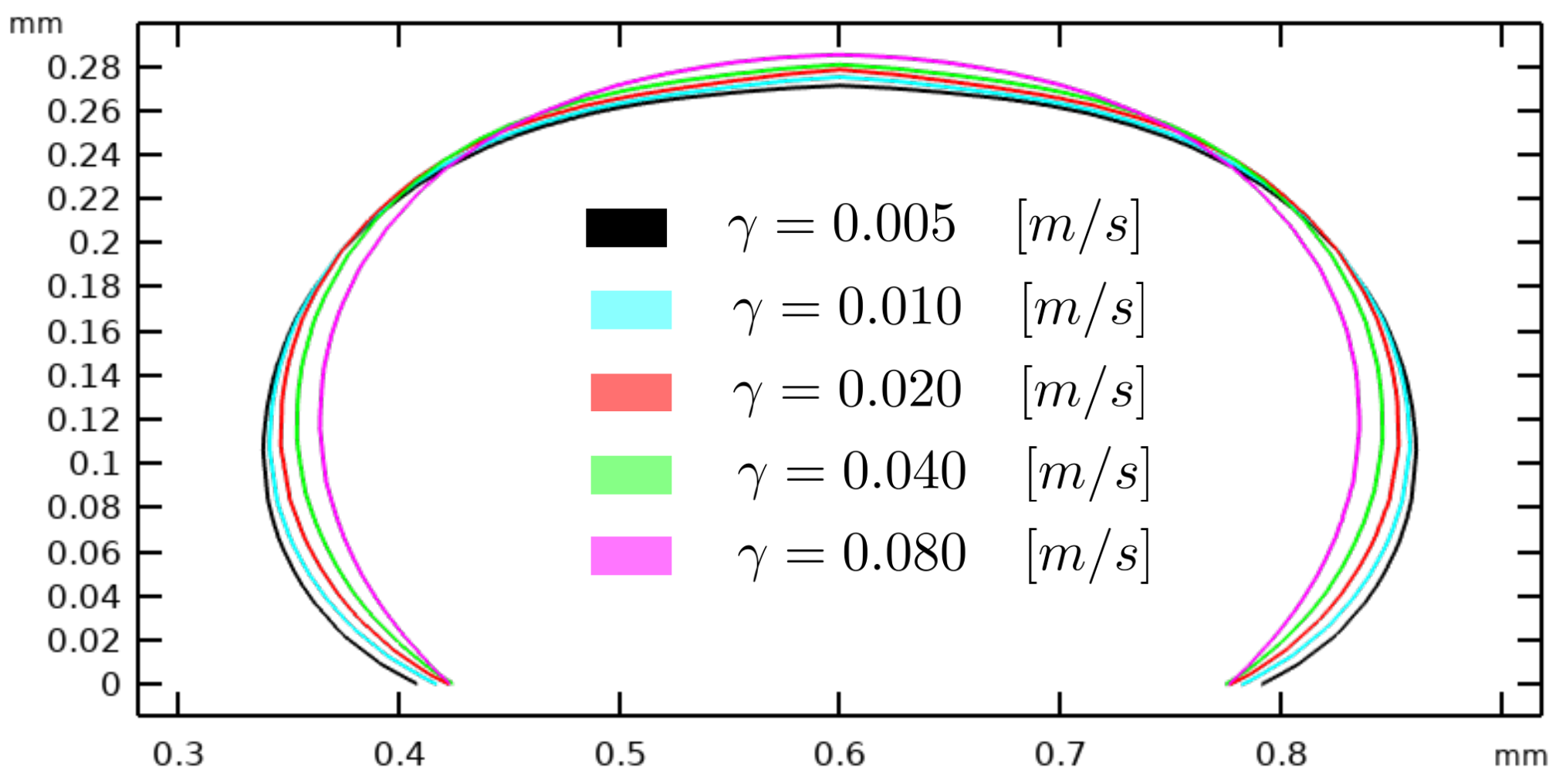}
  \caption{Effects of $\gamma$ on cross-sections}
  \label{fig:sub2}
\end{figure}
We simulated the process for different values of the reinitialization parameter ($\gamma$) to evaluate its effects on mass conservation. In Fig. 3, we see that the strand cross-section has an ellipsoidal shape and when $\gamma$ increases, the major axis shrinks, while the minor axis expands. The surface areas ($A_s$) obtained in the simulation for $\gamma$ values around the printing speed $\gamma_{ref} = 0.02 \ m/s$  are shown in Table 4. Decreasing $\gamma$ to $0.005\ m/s$ produced a slight improvement from around $5$ to $4 \%$ error. In contrast, increasing $\gamma$ had a greater impact on the precision of the results, as can be seen in the errors ($\Delta A$)  of the last two parameters $\gamma = \{0.04, 0.08\} \ m/s$  in Table 4. The simulation time was almost the same for the first two parameters compared to $\gamma_{ref}$. Furthermore, it was also noticed that for the last two parameters considered, there is a significant increase in the computation time.
\begin{table}
\caption{Effects of $\gamma$ on the strand cross-section}
\centering
\begin{tabular}{|c|c|c|c|}
\hline
\textbf{$\gamma \ [m/s]$} & \textbf{$A_s \times 1000 \ [mm^2]$} & \textbf{$\Delta A \%$} \\
\hline
0.005 & 60.34 & 3.96 \\
\hline
0.01 & 60.11 & 4.33 \\
\hline
0.02 & 59.65 & 5.07 \\
\hline
0.04 & 58.74 & 6.52  \\
\hline
0.08 & 57.06 & 9.19 \\
\hline
\end{tabular}
\label{tab:t1}
\end{table}

\subsubsection{Effects of the interface thickness}
Since the default interface thickness used in COMSOL considers the maximum element size, we analyzed the effects of reducing its value. The default interface thickness for Mesh 1 was $\epsilon_{ref} = 0.043\  [mm]$, which is approximately 7 times less than the nozzle-to-bed distance defined ($\Delta Z = 0.32 \ [mm]$). For this analysis, we considered changes based on the factor $\epsilon_f = \epsilon/\epsilon_{ref}$ and used $\gamma = 0.02 \ m/s$. Decreasing $\epsilon_f$ reduced the mass conservation error, as shown in Table 5. The parameters $\epsilon_f = \{0.9, 0.8\}$ approximately halve the error, and the following two parameters $\epsilon_f = \{0.7, 0.6 \}$ give an error below $1 \%$. Another interesting aspect of the results was that $A_s$ starts to exceed the reference value $A_f$ for the last two parameters, and there was an increase in the error when reducing $\epsilon_f$ from $0.6$ to $0.5$. The qualitative effects of $\epsilon$ on the cross-section are given in Fig. 4. Reducing the factor $\epsilon_f$ lead to ellipsoids with a larger major axis; however, the changes in the minor axis did not seem significant. Regarding the computational cost, most parameters resulted in a similar simulation time, except for the last, which took slightly longer than the others. 

\begin{table}
\caption{Effects of $\epsilon$ on the strand cross-section}
\centering
\begin{tabular}{|c|c|c|c|}
\hline
$ \epsilon_{f} \ [-]$ & $A_s \times 1000 \ [mm^2]$ & $\Delta A \%$\\
\hline
1.0 & 59.65 & 5.07 \\
\hline
0.9 & 60.89 & 3.10 \\
\hline
0.8 & 61.77 & 1.69 \\
\hline
0.7 & 62.67  &   0.68  \\
\hline
0.6 & 63.26  & 0.25  \\
\hline
0.5 & 63.74 & 1.44 \\
\hline
\end{tabular}
\label{tab:t2}
\end{table}
Based on the effects of the interface thickness and the influence of the mesh size on $A_s$, we also performed an additional test in which we decreased the mesh refinement and found a $\epsilon_f$ resulting in a low mass conservation error. The results of using a coarser mesh with $\epsilon_{ref} = 0.069 \ [mm]$ for different factors $\epsilon_f$ are given in Table 6. We observed a significant decrease in the cross-sectional error $\Delta A$ from $14$ to $0.2 \%$.
\begin{figure}[!htbp]
  \centering    \includegraphics[width=0.5\textwidth]{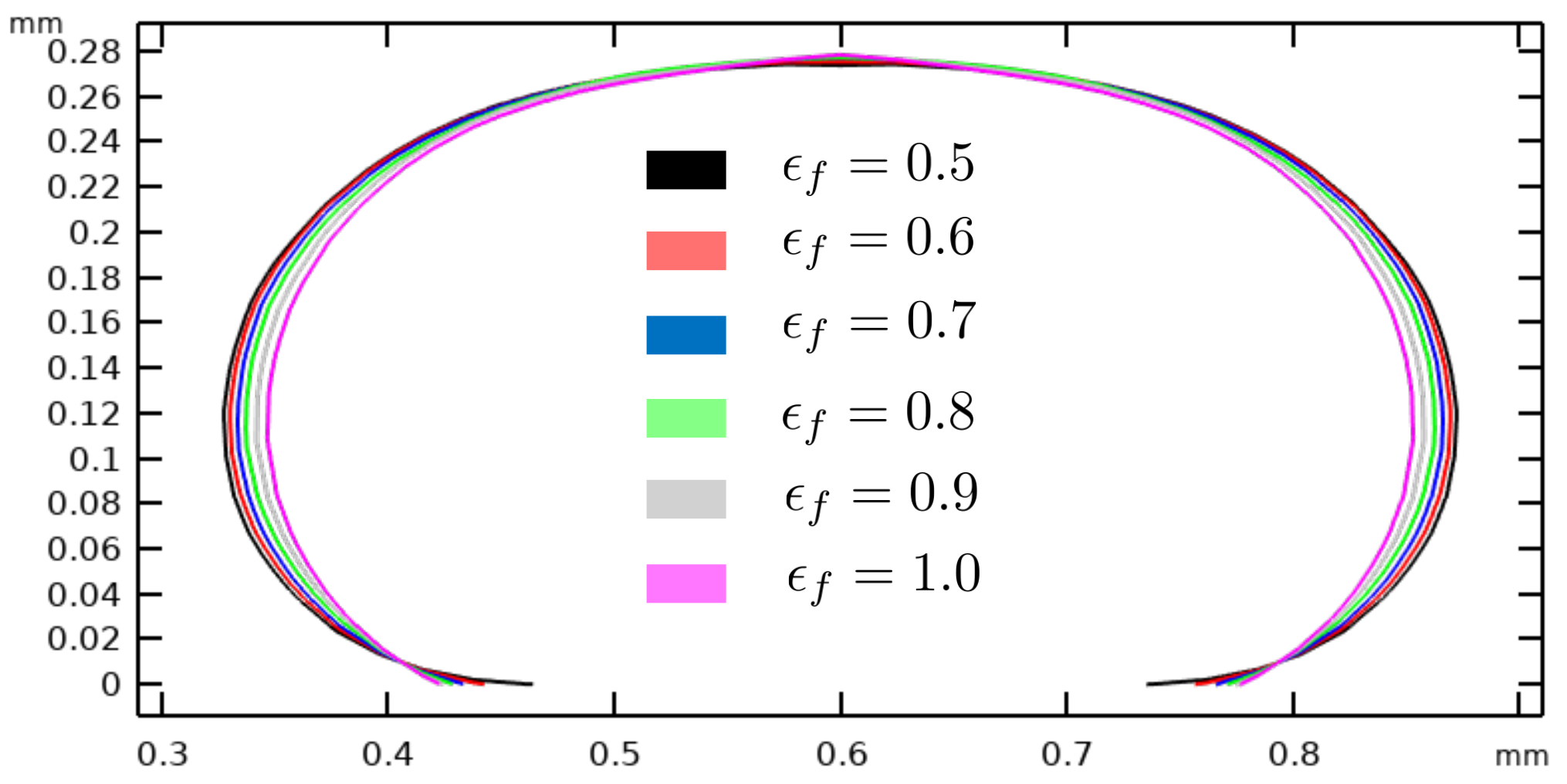}
  \caption{Effects of $\epsilon$ on cross-sections}
  \label{fig:sub3}
\end{figure}

The interface thickness for the best result obtained in Table 5 ($\epsilon_f = 0.6$) is $\epsilon = 0.0258 [mm]$, and the best result shown in Table 6 ($\epsilon_f = 0.4$) corresponds to $\epsilon = 0.0276 [mm]$. Therefore, what seems to be important is to have an adequate interface thickness around the fluid interface. In terms of the qualitative effects on the cross-section, we compared the two results in Fig. 5. We see that there are some slight differences in the ellipsoids, especially towards the top-centered part of the material. Finally, in terms of computational time, the simulation with the finer mesh and $\epsilon_f = 0.6$ took around 6h while the one with the coarser mesh and $\epsilon_f = 0.4$ took around 3h.

\begin{table}[h!]
\centering
\caption{$\epsilon$ refinement for a coarser mesh}
\begin{tabular}{|c|c|c|c|}
\hline
\textbf{$\epsilon_{f} \ [-]$} & \textbf{$A_s \times 1000 \ [mm^2]$} & \textbf{$\Delta A \%$} \\
\hline
0.8 & 53.65 & 14.61 \\
\hline
0.7 & 56.97 & 9.33 \\
\hline
0.6 & 59.57 & 5.19 \\
\hline
0.5 & 61.43  &   2.23  \\
\hline
0.4 & 62.68  & 0.24  \\
\hline
0.3 & 63.52 & 1.10 \\
\hline
\end{tabular}
\label{tab:t3}
\end{table}

\begin{figure}[!htbp]
  \centering    \includegraphics[width=0.5\textwidth]{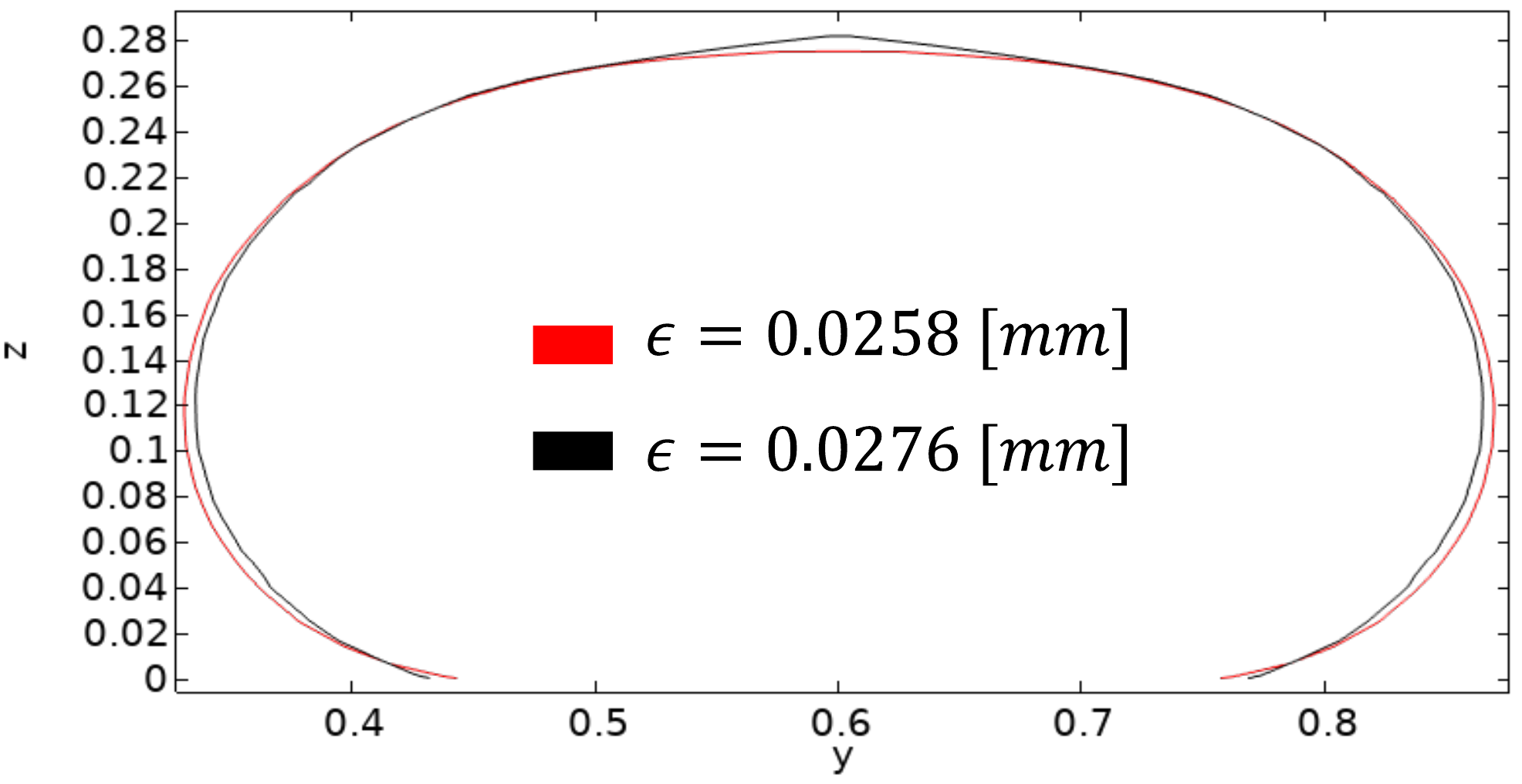}
  \caption{{Cross-sections for different mesh refinements with similar interface thickness}}
  \label{fig:sub4}
\end{figure}

\subsection{Mass conservation analysis for different conditions}
To conclude with the numerical analysis of single layer printing, we used the best level-set parameters found in the previous subsections ($\gamma = 0.005\,\text{m/s},\ \epsilon = 0.0276\,\text{mm}$) to evaluate the precision of the simulations for different nozzle-to-bed distances ($\Delta Z$) and printing speeds ($\bar{v}_x$). The reference areas were scaled according to the ratio $\bar{v}_x / v_p$ and the results are summarized in Table 7.
\begin{table}[h!]
\centering
\caption{Results for different printing parameters}
\begin{tabular}{|c|c|c|c|}
\hline
\textbf{$\Delta Z /D \ [-]$} &\textbf{$\bar{v}_x/v_p \ [-]$} & \textbf{$A_s \times 1000 \ [mm^2]$} & \textbf{$\Delta A \%$}  \\
\hline
0.6 & 0.5 & 123.71 & 0.48\\
\hline
0.6 & 0.75 & 83.70 & 0.09\\
\hline
0.6 & 1.0 & 62.53 & 1.55 \\
\hline
0.8 & 0.5 &   126.21 & 0.14  \\
\hline
0.8 & 0.75  & 84.36 & 0.70  \\
\hline
0.8 & 1.0 & 62.92 & 0.44 \\
\hline
1.0 & 0.5  &   127.85 & 0.88  \\
\hline
1.0 & 0.75 & 84.88 & 1.32 \\
\hline
1.0 & 1.0 & 62.28 & 1.74 \\
\hline
\end{tabular}
\label{tab:t4}
\end{table}
We see that independently of the printing parameters, the errors ($\Delta A$) are very low. In Figure 6, we present the cross-sectional areas obtained for these simulations. Although the reference areas $A_f$ did not change with $\Delta Z / D$, the major axis of the strand shrank when the nozzle-to-bed distance was increased. We also observed that the qualitative behavior of the results was good compared to the experimental results given in \cite{2_serdeczny_experimental_2018}. 
\begin{table}[h!]
\centering
\caption{Aspect ratio  comparison with $W/H - Sim$  and $ W/H - Exp$ in \cite{2_serdeczny_experimental_2018}}
\begin{tabular}{|c|c|c|c|c|}
\hline
\textbf{$\Delta Z /D \ [-]$} &\textbf{$\bar{v}_x/v_p \ [-]$} & \textbf{$W/H$} & \textbf{$W/H - Sim $}  & \textbf{$W/H - Exp $} \\
\hline
0.6 & 0.5 & 3.7 & 4.5 & 4.0\\
\hline
0.6 & 0.75 & 3.3 & 3.2 & 4.0\\
\hline
0.6 & 1.0 & 2.8 & 3.0 & 3.8\\
\hline
0.8 & 0.5 &  3.0 & 3.0 & 3.1 \\
\hline
0.8 & 0.75  & 2.4 & 2.4  & 2.9 \\
\hline
0.8 & 1.0 & 2.0 & 2.0 & 2.1 \\
\hline
1.0 & 0.5  &   2.6 & 2.6 & 2.5\\
\hline
1.0 & 0.75 & 1.9 & 1.8 & 2.1\\
\hline
1.0 & 1.0 & 1.5 & 1.5 & 1.8 \\
\hline
\end{tabular}
\label{tab:t4}
\end{table}
In Table 8, we present the aspect ratio $W/H$ obtained in this work and the values reported in \cite{2_serdeczny_experimental_2018}. There is a good agreement between the simulated (Sim) and experimental (Exp) data in \cite{2_serdeczny_experimental_2018} and the results obtained in this work for almost all the conditions evaluated. Furthermore, the shapes of the areas seem to match the experimental results more accurately than the areas simulated in \cite{2_serdeczny_experimental_2018}.

\begin{figure}[!htbp]
  \centering     \includegraphics[width=\textwidth]{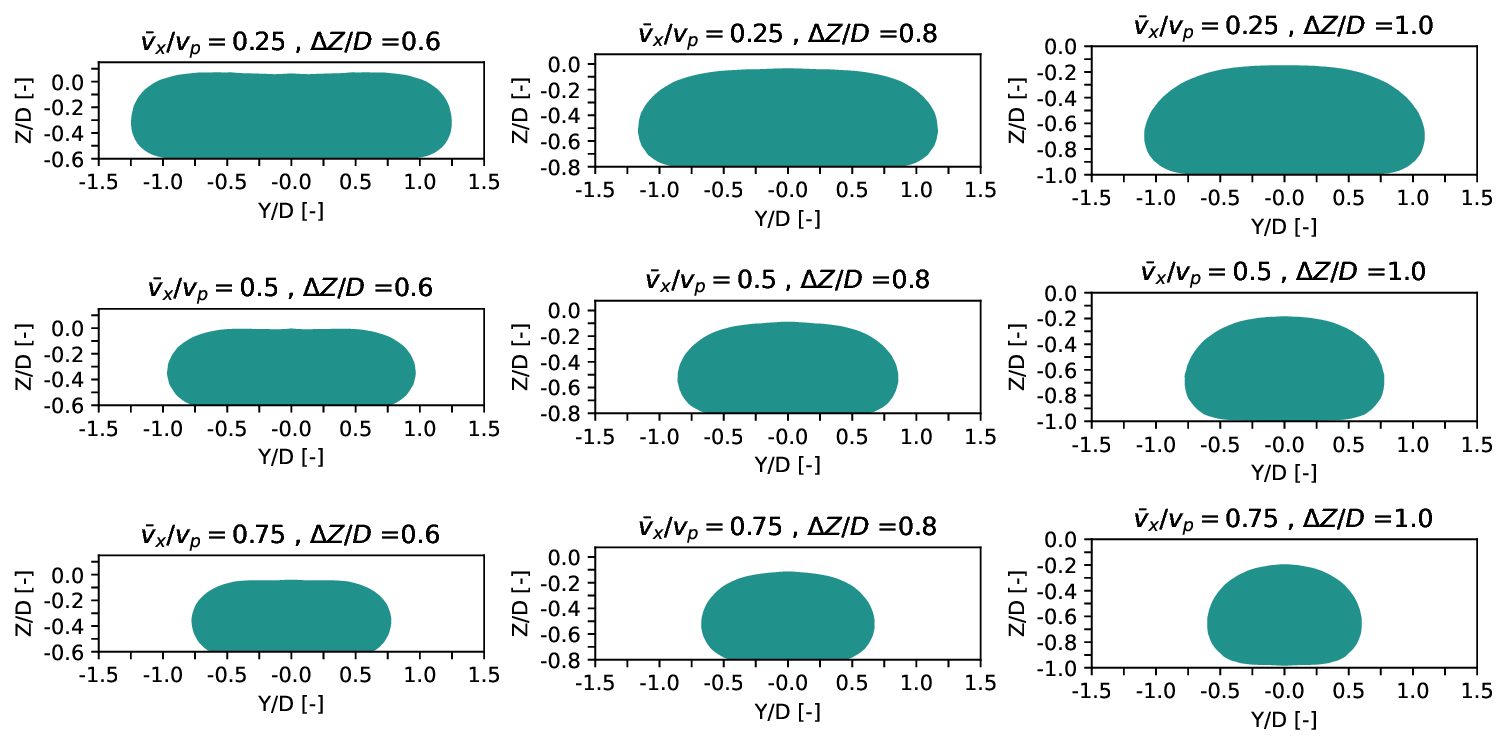}
  \caption{Cross-sectional shapes for different printing conditions}  \label{fig:sub7}
\end{figure}
{ 
\subsection{Two-layer printing of viscoplastic fluids}
In this subsection, we extend the previous analysis to evaluate the influence of the level-set parameters on multilayer printing involving varying fluid rheologies and nozzle geometries. In general, multilayer printing involves a complex interplay between layers, as a previously printed layer deforms when a new layer is deposited on top. Different printing strategies are commonly employed to achieve stable printing without excessive deformation. These include (i) wet-on-wet \citep{11_mollah_stability_2021} (ii) wet-on-semisolid \citep{431_mollah_computational_2023}, (iii) wet-on-solidifying \citep{431_mollah_virtual_2025}, and (iv) wet-on-solid \citep{11_mollah_stability_2021} approaches, particularly when printing viscoplastic materials such as thermosets, cementitious materials, and geopolymers. In this study, we focus on the wet-on-wet printing strategy, where the material rheology does not change during the printing process. Viscoplastic materials with different rheological properties are considered, together with nozzles featuring cylindrical and rectangular cross-sections. The analysis focuses on assessing how the level-set parameters affect strand morphology and overall mass conservation, and the results obtained with the conservative level-set formulation are compared against validated data from the literature.}
{\subsubsection{Concrete printing}
We selected the level-set parameters, $\epsilon$ and $\gamma$, following the insights gained from the results described in section 4.1. The reinitialization was defined as a fraction of the horizontal speed, $\gamma = 0.25|\bar{v}_x|$, and the interface thickness as a function of the maximum element size $\epsilon = 0.3m_{max}$. The maximum and minimum element sizes for the mesh were defined in terms of the diameter of the nozzle, $m_{min} =0.01D$ and $m_{max} =0.1D$. The other mesh settings were defined by the physics-based mesh feature in COMSOL.

To assess the accuracy of the simulation, we examined the strand cross sections for the first and second layers, as shown in Figure~7. The cross section of the first layer was obtained by defining a cut plane at $x = 200\,\text{mm}$ at $t = 9\,\text{s}$. For the second layer, we tracked the motion of this reference section from $t = 9\,\text{s}$ to $t = 20\,\text{s}$, which, based on the imposed printing speed, corresponded to a position of $x = 424.4\,\text{mm}$. According to the values in Table~2, the reference cross-sectional area for a single layer is $A_f = 550\,\text{mm}^2$, giving $A_f = 1100\,\text{mm}^2$ for two layers. 

In Figure~7, we compare the results obtained in this work (solid line) with the experimental measurements reported by \cite{10_spangenberg_numerical_2021}, as well as the numerical results obtained using the volume of fluid method (VOF) combined with a generalized Newtonian model, as presented in \cite{el_abbaoui_3d_2024}. At $t = 9\,\text{s}$, the contour plot shows a strand width of approximately $46\,\text{mm}$, which is approximately $5\,\text{mm}$ wider than the measurements and slightly larger than the VOF results with both no-slip (NS) and free slip (FS) boundary conditions at the nozzle wall. The simulated cross-sectional area was $A_s = 548\,\text{mm}^2$, corresponding to an error ($\Delta A$) of $0.36\%$. 
For the second layer, we observed a slight increase in the width of the first-layer strand to about $48\,\text{mm}$, along with minor changes in its contour. The upper strand exhibited a width of roughly $44\,\text{mm}$ and appeared less rounded than the bottom strand. The computed area for the two-layer configuration was $A_s = 1097\,\text{mm}^2$, yielding an error ($\Delta A$) of $0.27\%$.

 Overall, the two-layer cross-section obtained in this work showed good agreement with both the experimental data and the reference simulations. However, the interface between the two layers is smoothed when using the level-set method. As explained in the model description, with the level-set method, the material boundaries are defined by a diffuse and constant interface of finite thickness, and the properties of the phases are regularized with a Heaviside function that transitions smoothly between phases across multiple mesh elements. Consequently, when multiple layers are deposited sequentially, previously formed interfaces are continuously advected and reinitialized, resulting in numerical blending of layers rather than preservation of distinct material boundaries.
 
 Among the reference results, the VOF simulations with free-slip (FS) boundary conditions were closest to the experimental measurements. Despite the use of a simplified constitutive model and possible variations in the vertical motion profile between layers, the present results remain consistent with the reference studies \citep{10_spangenberg_numerical_2021, el_abbaoui_3d_2024} and with the analysis of mass-conservation.
 
\begin{figure}[!htbp]
  \centering     \includegraphics[width= \textwidth]{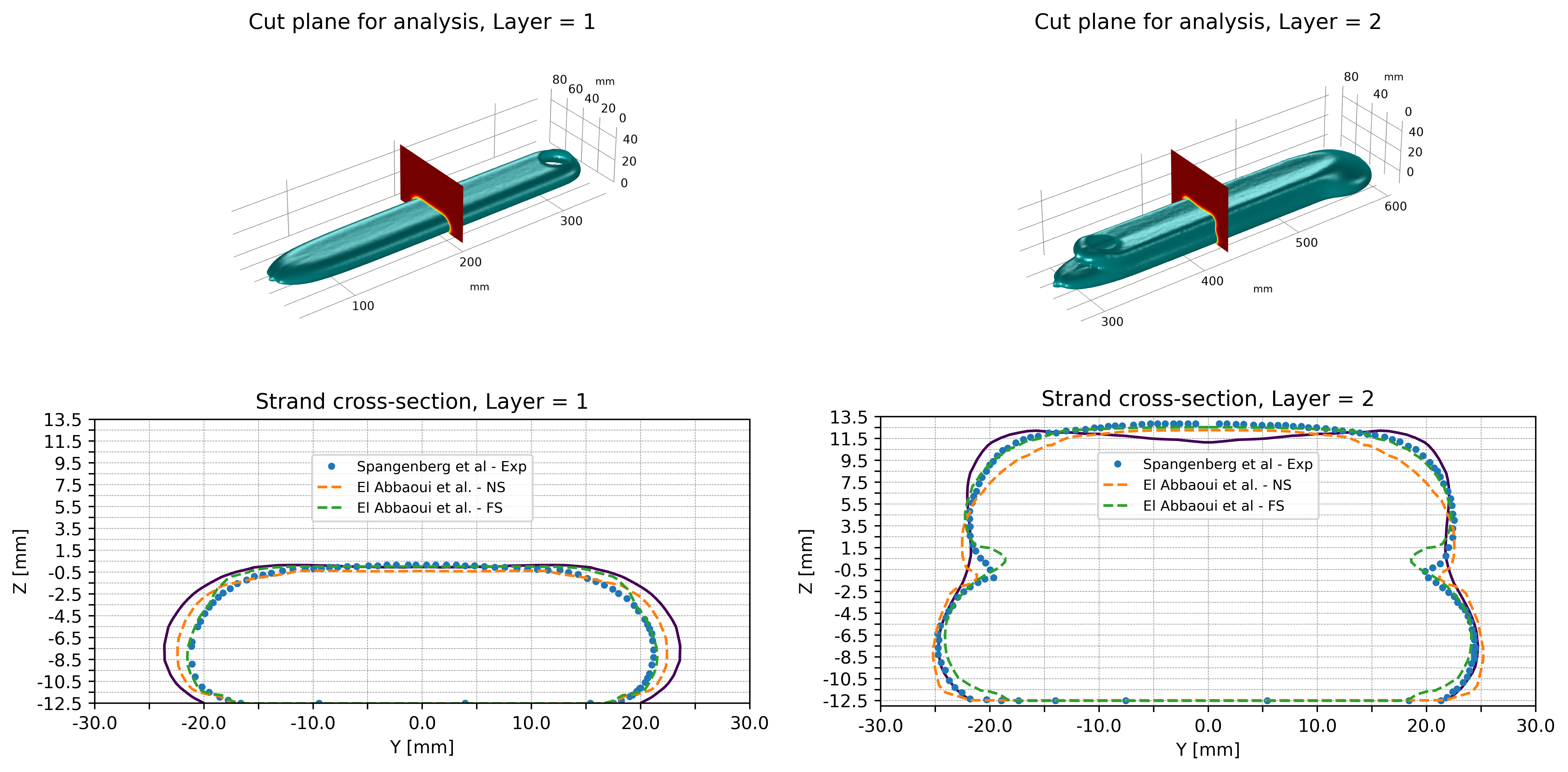}
  \caption{Cross-sectional shapes for multilayer concrete simulations}  \label{fig:sub8}
\end{figure}
\subsubsection{Metakaolin printing}
A similar analysis to the one described in the previous subsection was performed for the Metakaolin {printing}, in which we again evaluated the strand cross-sections for the first and second layers. The resulting cross-sections are shown in Figure 8. The profile of the first layer was obtained using a cutting plane at \(x=-20\ \mathrm{mm}\) at \(t=9\ \mathrm{s}\). For the second layer, we tracked the expected motion of this cutting plane according to the prescribed printing speed, reaching \(x=20\ \mathrm{mm}\) at \(t=22\ \mathrm{s}\). Based on the reference values listed in Table 2, the theoretical cross-sectional area for a single layer is \(A_f=36\ \mathrm{mm}^2\), hence \(A_f=72\ \mathrm{mm}^2\) for two layers. Finally, the simulated cross-sections were also compared with the experimental results reported in \cite{Mollah2025RapidCuringCFD}.

\begin{figure}[!htbp]
  \centering     \includegraphics[width=\textwidth]{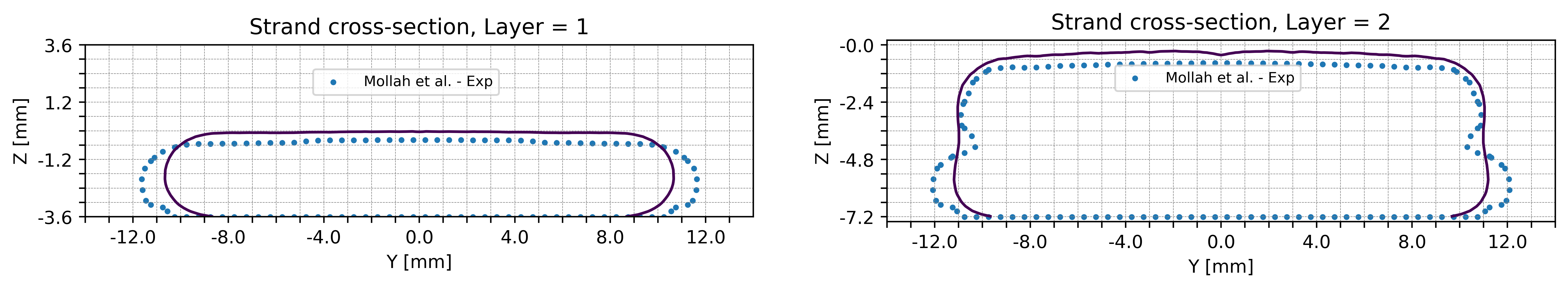}
  \caption{Cross-sectional shapes for multilayer Metakaolin simulations}  \label{fig:sub9}
\end{figure}
The contour plot for the first layer gives a strand width of \(w = 21.2\ \mathrm{mm}\), which is in good agreement with both the numerical results and the experimental measurements reported in \cite{Mollah2025RapidCuringCFD}, where the width was approximately \(w = 23\ \mathrm{mm}\). The simulated cross-sectional area for the first layer was \(A_s = 36.1\ \mathrm{mm}^2\), corresponding to an error of only 0.28\% relative to the reference area \(A_f\). Slight differences were observed in the width (1 mm), and the height (0.3 mm) when comparing the shape of the simulated (solid line) and measured cross-sections, shown in Figure 8. {The simulated cross-sectional area of the two layers was \(A_s = 73.85\ \mathrm{mm}^2\), corresponding to a deviation of 2.6\% from the reference area \(A_f\). Since the areas based on mass conservation are accurate, the discrepancies in first-layer width may be due to surface tension effects, the model regularization (Bingham-Papanastasiou), material slip on the substrate, or limitations in measurement resolution. Regarding the two-layer profile, the width of the material in the first layer increased slightly to approximately \(22.5\ \mathrm{mm}\), while the strand of the second layer was roughly \(22\ \mathrm{mm}\). The width of the first layer was therefore slightly lower than the experimental value of \(w = 24\ \mathrm{mm}\) reported in \cite{Mollah2025RapidCuringCFD}, while the width of the second layer closely matched the measurements reported. Furthermore, we also observed a smooth contrast between the two deposited layers, which is a consequence of the finite thickness interface representation of the conservative level-set method, as explained in the previous subsection.}

\section{Conclusion}\label{sec13}

{This work presented a systematic mass conservation analysis of MEX-AM simulations using the conservative level-set method. While the CLS formulation is attractive due to its smooth interface representation and numerical robustness, clear guidelines for selecting its key parameters have been lacking. By directly relating mass conservation errors to deviations in the steady-state cross-sectional area of printed strands, we provided a quantitative framework to assess the impact of the reinitialization parameter and interface thickness.

For single-layer deposition, the results show that mass conservation errors strongly depend on both the reinitialization parameter ($\gamma$) and the interface thickness ($\epsilon$). The results demonstrate that proper tuning of the parameters decreased mass conservation errors, leading to more accurate predictions of strand geometry. However, excessive reduction of the level-set parameters yielded diminishing improvements while increasing computational cost. Among the two parameters, the interface thickness was found to have a stronger influence on mass conservation accuracy than the reinitialization parameter, reducing the error below $1\%$. Proper selection of the interface thickness not only reduces steady-state errors but also enables the use of coarser meshes without significant compromises in accuracy, thereby alleviating the stringent mesh requirements and computational cost typically associated with level-set-based simulations. This finding highlights that interface thickness selection can be more effective than aggressive mesh refinement or reinitialization tuning in improving CLS-based simulation accuracy.

The effectiveness of the proposed parameter selection strategy was further validated for different printing speeds and nozzle-to-bed distances. For all tested cases of single-layer printing, mass conservation errors remained below $2\%$ Furthermore, the methodology was also successfully extended to multilayer simulations involving Bingham-type fluids and different nozzle geometries. In two-layer concrete and metakaolin printing simulations, the total cross-sectional area errors were below $0.5\%$ for concrete and below $3\%$ for metakaolin, while the predicted strand shapes showed good qualitative and quantitative agreement with experimentally validated results available in the literature.

Despite these encouraging results, the inherent limitations of the CLS formulation must be acknowledged. Due to the use of a diffuse interface with finite thickness and smooth Heaviside regularization of material properties, deposited layers are not preserved as sharp, distinct interfaces. In multilayer simulations, previously formed interfaces are smoothed, and there is partial blending between layers. While using a diffuse interface improves numerical stability and avoids spurious oscillations, it limits the ability of the CLS method to accurately represent sharp interlayer boundaries or material discontinuities that may be relevant for mechanical performance and interlayer bonding analysis.

Future work will focus on incorporating adaptive mesh refinement focused on the evolving interface, extending the analysis to viscoelastic and thixotropic material models, as well as integrating curing or solidification mechanisms, to further improve the predictive capability of the simulations. Finally, the quantitative mass conservation metrics developed in this work provide practical and transferable guidelines for tuning conservative level-set parameters in MEX-AM simulations and could be leveraged for automated parameter tuning.
}

\backmatter

\bibliography{sn-bibliography}% common bib file

@article{431_mollah_computational_2023,
	title = {Computational analysis of yield stress buildup and stability of deposited layers in material extrusion additive manufacturing},
	volume = {71},
	issn = {22148604},	
	language = {en},
	journal = {Additive Manufacturing},
	author = {Mollah, Md Tusher and Comminal, Raphaël and Serdeczny, Marcin P. and Šeta, Berin and Spangenberg, Jon},
	month = jun,
	year = {2023},
	pages = {103605},
}

@misc{431_mollah_virtual_2025,
	title = {A {Virtual} {3D} {Printing} {Framework} for {Off}-{Earth} {Construction}},
	copyright = {https://creativecommons.org/licenses/by/4.0/},	
	author = {Mollah, Md Tusher and Šeta, Berin and Spangenberg, Jon},
	month = jan,
	year = {2025},
}

@article{el_abbaoui_3d_2024,
	title = {{3D} concrete printing using computational fluid dynamics: {Modeling} of material extrusion with slip boundaries},
	volume = {118},
	issn = {15266125},
	shorttitle = {{3D} concrete printing using computational fluid dynamics},
	language = {en},
	urldate = {2026-01-06},
	journal = {Journal of Manufacturing Processes},
	author = {El Abbaoui, Khalid and Al Korachi, Issam and El Jai and Mostapha and Šeta, Berin and Mollah, Md. Tusher},
	month = may,
	year = {2024},
	pages = {448--459},
}

@article{serdeczny_viscoelastic_2022,
	title = {Viscoelastic simulation and optimisation of the polymer flow through the hot-end during filament-based material extrusion additive manufacturing},
	volume = {17},
	issn = {1745-2759, 1745-2767},
	language = {en},
	number = {2},
	urldate = {2025-12-23},
	journal = {Virtual and Physical Prototyping},
	author = {Serdeczny, Marcin P. and Comminal, Raphaël and Mollah, Md. Tusher and Pedersen, David B. and Spangenberg, Jon},
	month = apr,
	year = {2022},
	pages = {205--219},
}

@article{co_comminal_modelling_2020,
	title = {Modelling of {3D} concrete printing based on computational fluid dynamics},
	volume = {138},
	issn = {00088846},
	language = {en},
	urldate = {2025-09-03},
	journal = {Cement and Concrete Research},
	author = {Comminal, Raphael and Leal Da Silva, Wilson Ricardo and Andersen, Thomas Juul and Stang, Henrik and Spangenberg, Jon},
	month = dec,
	year = {2020},
	pages = {106256},
}

@inproceedings{Mollah2025RapidCuringCFD,
  author    = {Md. Tusher Mollah and Jacob Van Wonterghem and Hjalte Hedegaard Rasmussen 
               and Dan Meng and Berin {\v{S}}eta and Sahand Rahemipoor 
               and Daniel Helmuth Meile and Navid Ranjbar and Jon Spangenberg},
  title     = {Computational Fluid Dynamics Modeling and Experimental Analysis of Rapid Curing 
               in 3D Printing for Structural Applications},
  booktitle = {Proceedings of the 2025 Annual International Solid Freeform Fabrication Symposium (SFF Symposium 2025)},
  year      = {2025},
  address   = {Austin, Texas, USA}
}

@article{elgeti_deforming_2016,
	title = {Deforming {Fluid} {Domains} {Within} the {Finite} {Element} {Method}: {Five} {Mesh}-{Based} {Tracking} {Methods} in {Comparison}},
	volume = {23},
	issn = {1134-3060, 1886-1784},
	shorttitle = {Deforming {Fluid} {Domains} {Within} the {Finite} {Element} {Method}},
	language = {en},
	number = {2},
	urldate = {2025-09-03},
	journal = {Archives of Computational Methods in Engineering},
	author = {Elgeti, S. and Sauerland, H.},
	month = jun,
	year = {2016},
	pages = {323--361},
}

@article{wilms_formulation_2021,
	title = {Formulation engineering of food systems for {3D}-printing applications – {A} review},
	volume = {148},
	issn = {09639969},
	
	language = {en},
	urldate = {2022-08-09},
	journal = {Food Research International},
	author = {Wilms, P. and Daffner, K. and Kern, C. and Gras, S.L. and Schutyser, M.A.I. and Kohlus, R.},
	month = oct,
	year = {2021},
	pages = {110585},
}

@misc{comsol,
      author = "COMSOL, Inc.",
      title = "COMSOL Multiphysics Reference Manual, version 6.3",
      howpublished = "\url{www.comsol.com}",
      year = {2024}
    }

@inproceedings{20_Serdeczny2018_shear_thinning,
  author       = {Serdeczny, Marcin P. and Comminal, Raphaël and Pedersen, David B. and Spangenberg, Jon},
  title        = {Numerical study of the impact of shear thinning behavior on the strand deposition flow in the extrusion-based additive manufacturing},
  booktitle    = {Proceedings of the 18th International Conference of the European Society for Precision Engineering and Nanotechnology },
  editor       = {Billington, D. and Leach, R. K. and Phillips, D. and Riemer, O. and Savio, E.},
  location     = {Venice, Italy},
  month        = jun,
  year         = {2018},
  pages        = {283--284},
  isbn         = {9780995775121},
}

@article{5_Mirjalili2017_interface_capturing,
  author       = {Mirjalili, Shahab and Jain, Suhas S. and Dodd, Michael},
  title        = {Interface‑capturing methods for two‑phase flows: An overview and recent developments},
  journal      = {Center for Turbulence Research Annual Research Briefs},
  volume       = {2017},
  pages        = {117--135},
  year         = {2017},
}

@article{1_comminal_numerical_2018,
	title = {Numerical modeling of the strand deposition flow in extrusion-based additive manufacturing},
	volume = {20},
	issn = {22148604},
	language = {en},
	urldate = {2025-06-23},
	journal = {Additive Manufacturing},
	author = {Comminal, Raphaël and Serdeczny, Marcin P. and Pedersen, David B. and Spangenberg, Jon},
	month = mar,
	year = {2018},
	pages = {68--76},
}

@article{2_serdeczny_experimental_2018,
	title = {Experimental validation of a numerical model for the strand shape in material extrusion additive manufacturing},
	volume = {24},
	issn = {22148604},	
	language = {en},
	urldate = {2025-06-23},
	journal = {Additive Manufacturing},
	author = {Serdeczny, Marcin P. and Comminal, Raphaël and Pedersen, David B. and Spangenberg, Jon},
	month = dec,
	year = {2018},
	pages = {145--153},
}

@article{6_altiparmak_extrusion-based_2022,
	title = {Extrusion-based additive manufacturing technologies: {State} of the art and future perspectives},
	volume = {83},
	issn = {15266125},
	shorttitle = {Extrusion-based additive manufacturing technologies},
	language = {en},
	urldate = {2025-06-23},
	journal = {Journal of Manufacturing Processes},
	author = {Altıparmak, Sadettin Cem and Yardley, Victoria A. and Shi, Zhusheng and Lin, Jianguo},
	month = nov,
	year = {2022},
	pages = {607--636},
}

@article{7_naghieh_printabilitykey_2021,
	title = {Printability–{A} key issue in extrusion-based bioprinting},
	volume = {11},
	issn = {20951779},
	
	language = {en},
	number = {5},
	urldate = {2025-06-23},
	journal = {Journal of Pharmaceutical Analysis},
	author = {Naghieh, Saman and Chen, Xiongbiao},
	month = oct,
	year = {2021},
	pages = {564--579},
}

@article{8_bakrani_balani_investigation_2023,
	title = {An {Investigation} of the {Influence} of {Viscosity} and {Printing} {Parameters} on the {Extrudate} {Geometry} in the {Material} {Extrusion} {Process}},
	volume = {15},
	copyright = {https://creativecommons.org/licenses/by/4.0/},
	issn = {2073-4360},
	language = {en},
	number = {9},
	urldate = {2025-06-23},
	journal = {Polymers},
	author = {Bakrani Balani, Shahriar and Mokhtarian, Hossein and Salmi, Tiina and Coatanéa, Eric},
	month = may,
	year = {2023},
	pages = {2202},
}

@article{9_serdeczny_numerical_2019,
	title = {Numerical simulations of the mesostructure formation in material extrusion additive manufacturing},
	volume = {28},
	issn = {22148604},
	
	language = {en},
	urldate = {2025-06-23},
	journal = {Additive Manufacturing},
	author = {Serdeczny, Marcin P. and Comminal, Raphaël and Pedersen, David B. and Spangenberg, Jon},
	month = aug,
	year = {2019},
	pages = {419--429},
}

@article{10_spangenberg_numerical_2021,
	title = {Numerical simulation of multi-layer {3D} concrete printing},
	volume = {6},
	copyright = {https://creativecommons.org/licenses/by/4.0},
	issn = {2518-0231},
	urldate = {2025-06-23},
	journal = {RILEM Technical Letters},
	author = {Spangenberg, Jon and Leal Da Silva, Wilson Ricardo and Comminal, Raphaël and Mollah, Md. Tusher and Andersen, Thomas Juul and Stang, Henrik},
	month = oct,
	year = {2021},
	pages = {119--123},
}

@article{11_mollah_stability_2021,
	title = {Stability and deformations of deposited layers in material extrusion additive manufacturing},
	volume = {46},
	issn = {22148604},
	
	language = {en},
	urldate = {2025-06-23},
	journal = {Additive Manufacturing},
	author = {Mollah, Md Tusher and Comminal, Raphaël and Serdeczny, Marcin P. and Pedersen, David B. and Spangenberg, Jon},
	month = oct,
	year = {2021},
	pages = {102193},
}

@article{12_olsson_conservative_2005,
	title = {A conservative level set method for two phase flow},
	volume = {210},
	copyright = {https://www.elsevier.com/tdm/userlicense/1.0/},
	issn = {00219991},
	
	language = {en},
	number = {1},
	urldate = {2025-06-23},
	journal = {Journal of Computational Physics},
	author = {Olsson, Elin and Kreiss, Gunilla},
	month = nov,
	year = {2005},
	pages = {225--246},
}

@article{13_olsson_conservative_2007,
	title = {A conservative level set method for two phase flow {II}},
	volume = {225},
	copyright = {https://www.elsevier.com/tdm/userlicense/1.0/},
	issn = {00219991},
	
	language = {en},
	number = {1},
	urldate = {2025-06-23},
	journal = {Journal of Computational Physics},
	author = {Olsson, Elin and Kreiss, Gunilla and Zahedi, Sara},
	month = jul,
	year = {2007},
	pages = {785--807},
}

@article{14_mccaslin_localized_2014,
	title = {A localized re-initialization equation for the conservative level set method},
	volume = {262},
	issn = {00219991},
	
	language = {en},
	urldate = {2025-06-23},
	journal = {Journal of Computational Physics},
	author = {McCaslin, Jeremy O. and Desjardins, Olivier},
	month = apr,
	year = {2014},
	pages = {408--426},
}

@article{15_desjardins_accurate_2008,
	title = {An accurate conservative level set/ghost fluid method for simulating turbulent atomization},
	volume = {227},
	issn = {00219991},
	
	language = {en},
	number = {18},
	urldate = {2025-06-23},
	journal = {Journal of Computational Physics},
	author = {Desjardins, Olivier and Moureau, Vincent and Pitsch, Heinz},
	month = sep,
	year = {2008},
	pages = {8395--8416},
}

@article{16_chiodi_reformulation_2017,
	title = {A reformulation of the conservative level set reinitialization equation for accurate and robust simulation of complex multiphase flows},
	volume = {343},
	issn = {00219991},
	
	language = {en},
	urldate = {2025-06-23},
	journal = {Journal of Computational Physics},
	author = {Chiodi, Robert and Desjardins, Olivier},
	month = aug,
	year = {2017},
	pages = {186--200},
}

@article{17_jain_accurate_2022,
	title = {Accurate {Conservative} {Phase}-{Field} {Method} for {Simulation} of {Two}-{Phase} {Flows}},
	issn = {1556-5068},
	
	language = {en},
	urldate = {2025-06-23},
	journal = {SSRN Electronic Journal},
	author = {Jain, Suhas S.},
	year = {2022},
}

@article{172_jain_conservative_2020,
	title = {A conservative diffuse-interface method for compressible two-phase flows},
	volume = {418},
	issn = {00219991},	
	language = {en},
	urldate = {2025-10-23},
	journal = {Journal of Computational Physics},
	author = {Jain, Suhas S. and Mani, Ali and Moin, Parviz},
	month = oct,
	year = {2020},
	pages = {109606},
}

@article{18_chiu_conservative_2011,
	title = {A conservative phase field method for solving incompressible two-phase flows},
	volume = {230},
	copyright = {https://www.elsevier.com/tdm/userlicense/1.0/},
	issn = {00219991},
	
	language = {en},
	number = {1},
	urldate = {2025-06-23},
	journal = {Journal of Computational Physics},
	author = {Chiu, Pao-Hsiung and Lin, Yan-Ting},
	month = jan,
	year = {2011},
	pages = {185--204},
}

@article{19_mirjalili_conservative_2020,
	title = {A conservative diffuse interface method for two-phase flows with provable boundedness properties},
	volume = {401},
	issn = {00219991},
	
	language = {en},
	urldate = {2025-06-23},
	journal = {Journal of Computational Physics},
	author = {Mirjalili, Shahab and Ivey, Christopher B. and Mani, Ali},
	month = jan,
	year = {2020},
	pages = {109006},
}

@article{3_balta_numerical_2022,
	title = {Numerical and experimental analysis of bead cross-sectional geometry in fused filament fabrication},
	volume = {28},
	copyright = {https://www.emerald.com/insight/site-policies},
	issn = {1355-2546, 1355-2546},
	language = {en},
	number = {10},
	urldate = {2025-06-23},
	journal = {Rapid Prototyping Journal},
	author = {Balta, Efe C. and Altınkaynak, Atakan},
	month = oct,
	year = {2022},
	pages = {1882--1894},
}

@article{Lukhi2025,
  author    = {Mehul Lukhi and Christoph Mittermeier and Josef Kiendl},
  title     = {Multi-physics simulation of a material extrusion-based additive manufacturing process: towards understanding stress formation in the printed strand},
  journal   = {Progress in Additive Manufacturing},
  year      = {2025},
  volume    = {10},
  number    = {9},
  pages     = {6839--6853},
}

@article{BakraniBalani2025,
  author    = {Shahriar Bakrani Balani and Hossein Mokhtarian and Tiina Salmi and Eric Coatanéa},
  title     = {Numerical simulation of the shear rate in the fused filament fabrication process of poly-ether-ether-ketone (PEEK)},
  journal   = {Progress in Additive Manufacturing},
  year      = {2025},
  volume    = {10},
  number    = {8},
  pages     = {5057--5077},
}

@article{Mishra2022,
  author    = {Ases Akas Mishra and Affaf Momin and Matteo Strano and Kedarnath Rane},
  title     = {Implementation of viscosity and density models for improved numerical analysis of melt flow dynamics in the nozzle during extrusion-based additive manufacturing},
  journal   = {Progress in Additive Manufacturing},
  year      = {2022},
  volume    = {7},
  number    = {1},
  pages     = {41--54},
}

\end{document}